\documentstyle[aps,prb,epsfig,amsmath,amssymb]{revtex}
\begin{document}
\draft

\title{Unpaired and spin-singlet paired states of a two-dimensional
electron gas in a perpendicular magnetic field}
\author{M. Polini$^{1}$, K. Moulopoulos$^{2}$, B. Davoudi$^{1,3}$ and
M. P. Tosi$^{1}$}
\address{$^{1}$NEST-INFM and Classe di Scienze, Scuola Normale Superiore, I-56126 Pisa, Italy\\
$^{2}$Department of Physics, University of Cyprus, P.O. Box 20537, 1678
Nicosia, Cyprus\\ 
$^{3}$Institute for Studies in Theoretical Physics and Mathematics, Tehran
19395-5531, Iran
}
\maketitle

\vspace{0.5 cm}
\begin{abstract}
We present a variational study of both unpaired and spin-singlet
paired states induced in a two-dimensional electron gas at low density
by a perpendicular magnetic field. It is based on an improved circular-cell
approximation which leads to a number of closed analytical
results. The ground-state energy of the Wigner crystal containing a
single electron per cell 
in the lowest Landau level is obtained as a function of the
filling factor $\nu$: the results are in good agreement  with those of
earlier approaches and predict $\nu_{c} \approx 0.25$ for the upper
filling factor at which the solid-liquid transition occurs. 
A novel localized state of spin-singlet electron pairs is 
examined and found to be
a competitor of the unpaired state for filling factor $\nu >1$.
The corresponding phase boundary is quantitatively displayed 
in the magnetic field-electron density plane.    
\end{abstract}
\section{Introduction}\label{intro}
The nature of the ground-state of a two-dimensional (2D) many-electron 
system in a 
perpendicular magnetic field has been for a number of years a subject
of intense experimental and theoretical investigation. Since the
discovery of the Integer and Fractional Quantum Hall
Effects\cite{prange}, this has been one of the richest sources of
fundamental new physics in Condensed Matter. The transition from a
liquid to a Wigner crystal (WC) state, originating from strong
electron-electron correlations at low density in strong magnetic
fields, has received considerable attention (see {\it e.g.} Ref. 2 and
references given therein). More recently, novel and intriguing
behaviors of 2D electron transport in high Landau levels have been
reported\cite{jim}. In such regime of weak magnetic fields Koulakov
{\it et al.}\cite{KFS} predicted that charge density waves (CDW) would
break the translational symmetry in one direction and form
stripes. They also predicted stabilization of a novel 2D-CDW state
with more that one electron per unit cell. Such a ``bubble'' phase
arising from spin-triplet electron pair formation and 2D lattice
ordering has been confirmed in systematical numerical studies of
Haldane {\it et al.}\cite{haldane} and of Shibata and Yoshioka\cite{shibata}.

In the present work we introduce a new possible low-temperature phase
for a 2D electron system in a perpendicular magnetic field, namely a
localized paired-electron state consisting of a spin-singlet electron
pair in each cell, and investigate its stability at low densities
against the more conventional WC state containing a single electron
per cell. We have in mind situations where the Zeeman splitting can be neglected\cite{girless}. In free space, where the Land\`e $g$-factor of an electron is $g=2$, the Zeeman splitting is exactly equal to the cyclotron splitting. It turns out, however, that this is incorrect, for example, in a GaAs heterostructure for the following reasons: (i) the small effective mass in the conduction band increases the cyclotron energy by a factor of $m/m^* \sim 14$; (ii) the effective coupling of the electron spin to the external magnetic field is reduced by spin-orbit scattering by a factor of $-5$, making the effective Land\`e factor $g \sim -0.4$. Thus the Zeeman energy is about $70$ times smaller than the cyclotron energy\cite{girless}. 

This study has two main
motivations. Firstly, from the solution of the simple problem of
two-electrons moving on a plane under parabolic confinement\cite{taut}
it emerges that the effective radial potential for the relative motion
develops a pronounced minimum at sufficiently weak field. The minimum 
occurs even in the case of
vanishing angular momentum ($m=0$) and is the result of the
competition between the Coulomb repulsion and the localizing effects of
the magnetic field and of the confinement. This feature may hide, in
the many-body context  and at sufficiently low electron densities,
some new phase transition to a spin-singlet (for  $m=0$) paired
electron state in weak magnetic fields. Secondly, and following early
work on a $m=0$ paired state in the 3D electron gas\cite{russian},
such a state was found in a variational calculation\cite{ma} to be a
possible competitor to the conventional 3D-WC in zero field. A mixed
spin state with a preference for spin pairing has been proposed to
occur in some regions of the liquid phase from experiments on 2D
electron systems in the ultra-quantum limit\cite{experimental}.

In our evaluation of the energetics of the paired state we introduce
an improved circular-cell approximation for an interacting 2D electron
gas on a uniform positive background. We first study by this approach
the conventional unpaired-electron WC state in a magnetic
field\cite{mou}. Each electron is treated as a distributed charge
cloud, which can be taken to a good approximation to be localized
within  a cell if the density is not too high and the field is not too
small -- the radius of the circular cell being available  as an
additional variational parameter\cite{nagy}. Our results 
are in close agreement with those obtained in earlier
approaches. We then adopt a similar method to deal with two
interacting electrons within each circular cell. This yields new and
more complex expressions leading to a window of
thermodynamic stability  for the paired-electron state at higher
values of the electron density.  

The layout of the paper is briefly as follows. In Section \ref{sec2} we
present our treatment of the energy of the unpaired-electron 2D-WC in
a magnetic field and compare our results with earlier ones and with
the energy of the Laughlin liquid state. Section \ref{sec3} summarizes the
main results for the problem of two electrons moving in a plane under
an applied magnetic field. Section \ref{sec4} reports our treatment of the
singlet-paired state and Section \ref{sec5} gives a variational calculation of
the energy of this state and a determination of the phase boundary
between the unpaired and the singlet-paired states. Finally, Section
VI summarizes our main conclusions. Some technical points of detail
are worked out in an Appendix.
\section{Distributed-charge approach to the Wigner Crystal in a
magnetic field}\label{sec2}

A 2D system of carriers in a pure semiconductor sample subject to a
magnetic field $B$  is expected to undergo a transition to an ordered
triangular lattice structure at sufficiently low temperature $T$ and
filling factor $\nu$\cite{loz}. Here, $\nu= 2\pi\,n_{s} \ell^{2}$
where $n_{s}$ is the areal carrier density and 
$\ell=(\hbar c /eB)^{1/2}$ is the magnetic length. The
ground-state energy of such a 2D-WC has been evaluated by a number of
authors with various methods. In the following we shall indicate by
$\delta \epsilon$ the values of the ground-state energy after
subtraction of the energy of the lowest Landau Level (LL). 

The leading term of a low-density expansion is given by a classical
Madelung-potential calculation\cite{bonsa}, leading to 
$\delta \epsilon_{\rm \scriptstyle class}=-2.212206\, r^{-1}_{s}\,\,
Ryd^{*}= -0.782133\,\nu^{1/2}\,\,(e^{2}/\epsilon\,\ell)$ where
$\epsilon$ is the dielectric constant of the host medium and
$r_{s}a^{*}_{B}=(\pi n_{s})^{-1/2}$ with $Ryd^{*}$ and $a^{*}_{B}$ the
effective Rydberg and Bohr radius. Following early Hartree-Fock
calculations\cite{yoshi} and a classical-plasma simulation for the
determination of the energy of the Laughlin liquid\cite{mac}, a
variational calculation based on a correlated magnetophonon
wave-function for the WC\cite{girvin} gave the result
\begin{equation}\label{CWC}
\delta \epsilon_{\rm \scriptscriptstyle
CWC}=-0.782133\,\nu^{1/2}+0.2410\,\nu^{3/2}+0.16\,\nu^{5/2}
\,\,(e^{2}/\epsilon\,\ell)  
\end{equation}
and predicted that the crystal energy would be lower than that of the
liquid only at values of $\nu$ lower than $\nu_{c} \approx
0.14$. Subsequent work by Vignale\cite{vignale} used Current-Density
Functional Theory in the projected local-density approximation and
located the upper filling factor for the liquid-solid transition
at $\nu_{c} \approx 0.25$, in fairly good agreement with the
experimental evidence.

In our approach we model the 2D interacting electron system on a
uniform positive background as a collection of non-interacting disks,
each having a Wigner-Seitz radius $r_{\rm \scriptscriptstyle WS}$ to
be eventually treated as a variational parameter and containing a
single electron in a trial wave function with a variational width
parameter $\sigma$. Specifically, we adopt a Gaussian trial wave
function normalized to unity, 
\begin{equation}\label{prima}
\phi_{\rm \scriptstyle SP}(r)=(\pi \sigma^2)^{-1/2}\,\, 
\exp{(-r^{2}/2  \sigma^{2})}
\end{equation}
The single-particle density thus is 
$\rho(r)=\phi^{2}_{\rm \scriptstyle SP}(r)$.
 Taking the areal density of the background as
$\rho_{b}(r)=n_{s}\theta(r_{\rm \scriptscriptstyle WS}-r)$, the total
electronic charge inside the disk radius is determined by 
\begin{equation}\label{lecca}
\int\! d^{2}r \,\phi^{2}_{\rm
\scriptstyle SP}(r)\,\theta(r_{\rm \scriptscriptstyle WS}-r)
=1-\exp{(-r_{\rm \scriptscriptstyle WS}^{2}/\sigma^2)} 
\end{equation}
and we must require $\sigma \ll r_{\rm \scriptscriptstyle WS}$ in order
to avoid charge leakage errors. As noted by
Nagy\cite{nagy}, the handling of $r_{\rm \scriptscriptstyle WS}$ as a
variational parameter ensures that the error involved in setting the
electrical potential outside the disk to
zero is minimized.
\subsection{Electron-background and background-background interaction energy}
The potential energy $V_{\mbox{\tiny
in}}(r)$ created by the positive background inside the circular cell
is given in terms of the Gauss hypergeometric function ${\rm F}(a,b;c;d)$
by\cite{nagy}
\begin{equation}\label{non}
V_{\mbox{\tiny 
in}}(r)=\frac{4}{r_{s}}\, 
{\rm F}\left(\frac{1}{2}, -\frac{1}{2}; 1; r^2/r^2_{\rm
\scriptscriptstyle WS}\right)\,\,\frac{r_{\rm \scriptscriptstyle
WS}}{r_{s}\,a^{*}_{B}}\,\, Ryd^{*}.  
\end{equation}
This result is equivalent to the more familiar expression 
\begin{equation}\label{vale}
V_{\mbox{\tiny in}}(r)=\frac{8}{\pi\,r_{s}}
\,{\rm E}(r^{2}/r^{2}_{\rm \scriptscriptstyle
WS})\,\,\frac{r_{\rm \scriptscriptstyle WS}}{r_{s}\, 
a^{*}_{B}}\,\, Ryd^{*}  
\end{equation}
where ${\rm E}(x)$ is the complete elliptic integral of the first kind
(see Ref. \onlinecite{abram}, p. 591).

The self-interaction energy of the background is thus given by 
\begin{equation}\label{ture}
\epsilon_{bb}(r_{s}, r_{\rm \scriptscriptstyle WS})= \frac{1}{2}
\int\!d^{2}r\, \rho_{b}(r)\,V_{\mbox{\tiny in}}(r)
=\frac{16}{3\pi r_{s}}\,
\left(\frac{r_{\rm \scriptscriptstyle WS}}{r_{s}a^{*}_{B}}\right)^3\,
\,Ryd^{*}\, ,
\end{equation}
having used the result $\int_{0}^{1}\,E(x^{2})\,
x\,dx=2/3$. Similarly, the interaction energy of the electron with the
background is 
\begin{eqnarray}\label{first}
\epsilon_{eb}(r_{s}, r_{\rm \scriptscriptstyle WS}, \sigma)&\!=\!& 2 \pi
\int_{0}^{r_{\rm \scriptscriptstyle WS}}\!r \,dr \,\phi^{2}_{\rm
\scriptstyle SP}(r) \,V_{\mbox{\tiny in}}(r)\nonumber \\
&\!=\!& -\frac{16
\,r_{\rm \scriptscriptstyle WS}}{\pi r^2_{s} 
a^{*}_{B}}\,\int_{0}^{r_{\rm \scriptscriptstyle WS}/\sigma}\!\!x\,{\rm
E}\left(\sigma^2 \,x^2/r^2_{\rm \scriptscriptstyle WS}\right)\,
e^{-x^2}\,dx   \,\,\,Ryd^{*}.
\end{eqnarray}
For $\sigma \ll r_{\rm \scriptscriptstyle WS}$ Eq. (\ref{first}) can
be replaced by the approximate analytic expression
\begin{equation}\label{pippo}
\epsilon_{eb}(r_{s}, r_{\rm \scriptscriptstyle WS}, \sigma)=
\left[-\frac{4}{r_{s}}\,\frac{r_{\rm \scriptscriptstyle WS}}{r_{s}
a^{*}_{B}}+\frac{1}{r^3_{s}}\,\frac{(\sigma/a^{*}_{B})^2}{(r_{\rm
\scriptscriptstyle WS}/r_{s}a^{*}_{B})} 
\right]\, Ryd^{*}.
\end{equation}
This so-called harmonic approximation (HA) for the electron-background
energy is obtained from Eq. (\ref{first}) by extending the range of
integration up to $\infty$ and by using the expansion of the elliptic
integral ${\rm E}(x)= (\pi/2)(1-x/4)+o(x^2)$ (see \onlinecite{abram},
p. 591). 

An alternative way to calculate the electron-background energy is by
using the expression 
\begin{equation}\label{complete}
\epsilon_{eb}(r_{s}, r_{\rm \scriptscriptstyle WS}, \sigma)= 2 \pi
\int_{0}^{r_{\rm \scriptscriptstyle WS}}\rho_{b}(r)\, 
V_{\rm \scriptstyle SP}(r) \,r \,dr\, , 
\end{equation}
where $V_{\rm \scriptstyle SP}(r)$ is the electrical potential created
by a single electron in the state (\ref{prima}). This can be obtained
in closed form,  
\begin{equation}\label{cmm}
V_{\rm \scriptstyle SP}(r)=-\frac{2
\sqrt{\pi}}{(\sigma/a^{*}_{B})}\,\exp{(-r^{2}/2\sigma^{2})}\,
{\rm I}_{0}(r^{2}/2\sigma^{2})\,\,Ryd^{*}
\end{equation}
where ${\rm I}_{n}(x)$ is the modified Bessel
function of the $n$-th order (see Appendix). The evaluation of
Eq. (\ref{complete}) can then be carried out in closed form, with the
result  
\begin{equation}\label{cm}
\epsilon_{eb}(r_{s}, r_{\rm \scriptscriptstyle WS},
\sigma)=-\frac{2\sqrt{\pi}}{(\sigma/ 
a^{*}_{B})}\,\left(\frac{r_{\rm \scriptscriptstyle
WS}}{r_{s}a^{*}_{B}}\right)^2
\,\exp{(-r_{\rm \scriptscriptstyle WS}^{2}/2 \sigma^2)}\left[
{\rm I}_{0}(r_{\rm \scriptscriptstyle WS}^{2}/2 \sigma^2)+{\rm
I}_{1}(r_{\rm \scriptscriptstyle WS}^{2}/2 
\sigma^2)\right]\,\, Ryd^{*} 
\end{equation} 
(see Appendix). The expression
(\ref{pippo}) is recovered from Eq. (\ref{cm}) for $r_{\rm
\scriptscriptstyle WS} \gg \sigma$ by using the asymptotic
expansion of the Bessel functions (see \onlinecite{abram}, p. 377). We shall see
below that the ``anharmonic corrections'' implied by Eq. (\ref{cm})
are crucial in a comparison with earlier calculations. 
\subsection{Total energy and variational procedure}
The total energy per electron in a 
perpendicular magnetic field ${\bf B}$ is 
\begin{equation}\label{tot}
\epsilon_{t}=\epsilon_{k}(\sigma, B)+ 
\epsilon_{eb}(r_{s}, r_{\rm \scriptscriptstyle WS},
\sigma)+\epsilon_{bb}(r_{s}, r_{\rm \scriptscriptstyle WS})
\end{equation}
where $\epsilon_{k}(\sigma, B)$ is the kinetic energy,
\begin{eqnarray}\label{13}
\epsilon_{k}(\sigma, B)&\!=\!&\int\!d^{2} r\, 
\phi_{\rm \scriptstyle SP}(r)\frac{1}{2m^{*}}
\left({\bf p}+\frac{e}{c}{\bf A}\right)^{2}\phi_{\rm \scriptstyle
SP}(r) \nonumber \\
&\!=\!& 2\pi\, \int_{0}^{+\infty} r\,
\phi_{\rm \scriptstyle SP}(r)\left[-\frac{\hbar^{2}}{2m^{*}\,r}\, 
\partial_{r}(r \partial_{r}) +\frac{1}{2} m^{*}\omega_{0}^{2} r^{2}\, 
\right]\phi_{\rm \scriptstyle SP}(r) \,  d r   
\end{eqnarray}
with $\omega_{0}=eB/2m^{*}c$. We have taken the vector potential in
the symmetric gauge, ${\bf A}={\bf B}\times
{\bf r}/2$, and used the fact that the wave function (\ref{prima}) is
an isotropic state of zero angular momentum. Eq. (\ref{13}) yields
\begin{equation}\label{megane}
\epsilon_{k}(\sigma, B)=\left[\frac{1}{(\sigma/a^{*}_{B})^2} + \lambda_{B}\,
\left(\frac{\sigma}{a^{*}_{B}}\right)^{2}\right]\,\, Ryd^{*}  
\end{equation}
where $\lambda_{B}\equiv m^{*}\omega_{0}^{2}
a^{*\,\,2}_{B}/(2\,\,Ryd^{*})\simeq 3.6 \cdot
10^{-12}\,\,\left(\epsilon\,m/m^{*}\right)^{4}\,\,(B/{\rm
Tesla})^{2}$. For a GaAs heterostructure 
we have $\lambda_{B}\simeq 0.4$ if $B=10$ Tesla.

\begin{figure}[h!]
\centerline{\mbox{\psfig{figure=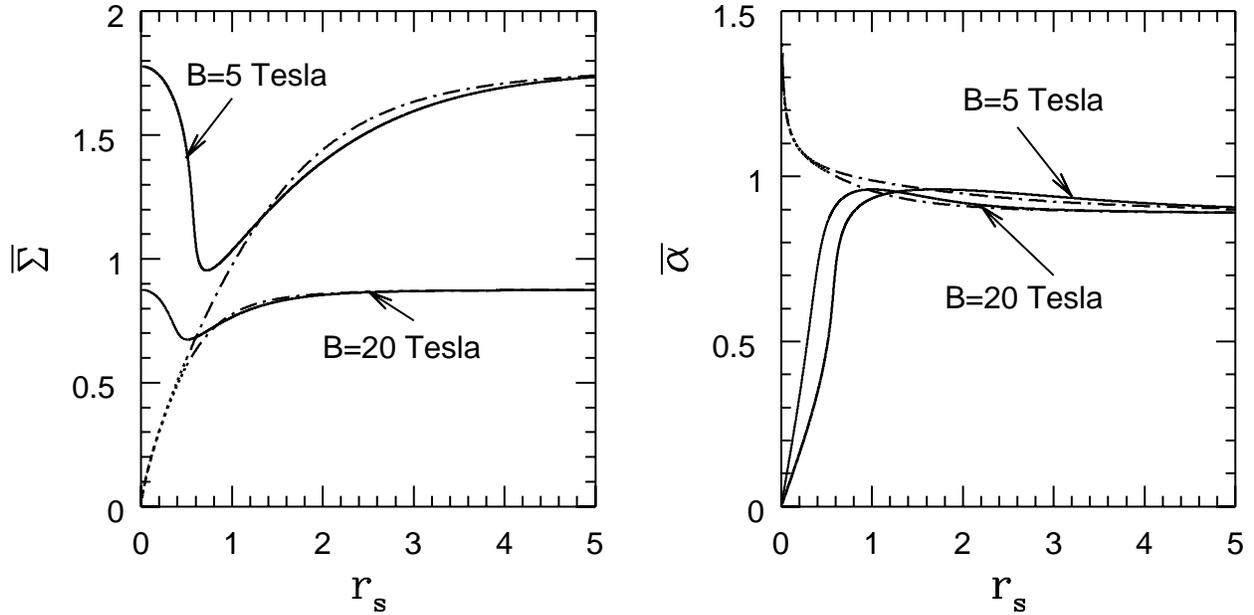, angle =0, width = 18 cm}}}
\caption{Equilibrium values of the reduced width ${\bar \Sigma}\equiv
(\sigma/a^{*}_{B})_{\rm \scriptscriptstyle eq}$ (left panel) and of
the reduced disk radius ${\bar \alpha}\equiv (r_{\rm
\scriptscriptstyle WS}/r_{s}a^{*}_{B})_{\rm \scriptscriptstyle eq}$
(right panel) as functions of $r_{s}$ and for two values of the 
magnetic field. The asymptotic values at large $r_{s}$ are
$\sigma_{\rm \scriptscriptstyle
eq}=(\hbar/m^{*}\omega_{0})^{1/2}$ and ${\bar \alpha}=\sqrt{\pi}/2$,
the latter being the value obtained by Nagy\cite{nagy} in zero
field. The dot-dashed lines are obtained from the harmonic
approximation, which is becoming approximately valid for $r_{s}\geq 1$.}
\label{fig1}
\end{figure}

\begin{figure}[h!]
\centerline{\mbox{\psfig{figure=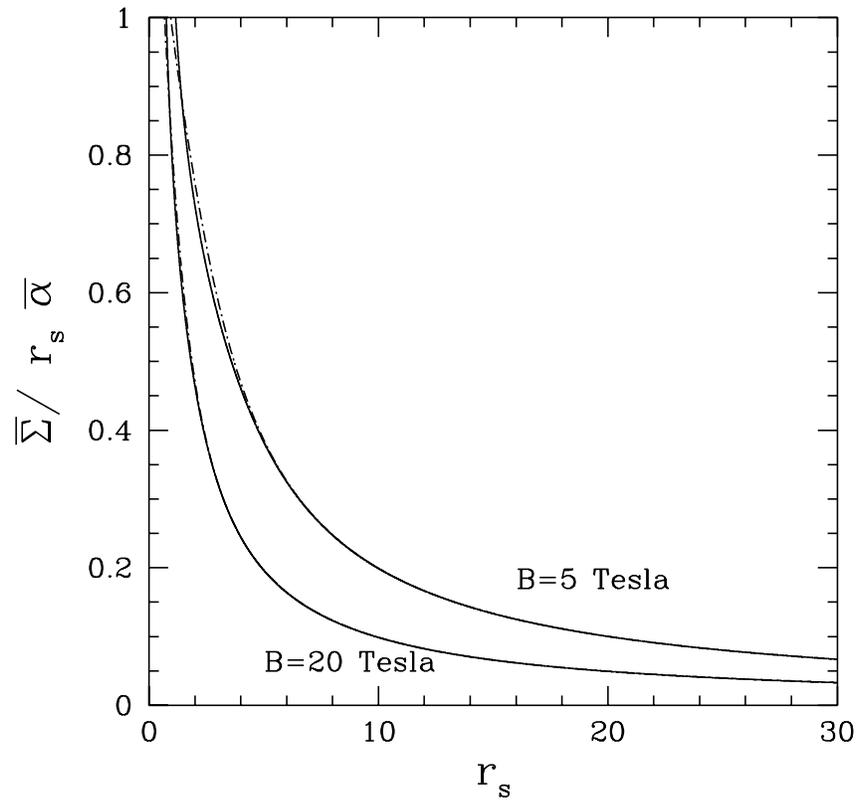, angle =0, width = 12 cm}}}
\caption{Consistency ratio ${\bar \Sigma}/(r_{s} {\bar \alpha})$ as a
function of $r_{s}$ for two values of the magnetic field. The
dot-dashed lines are obtained from the harmonic approximation.}
\label{fig2}
\end{figure}

Equation (\ref{tot}), after insertion of Eqs. (\ref{ture}), (\ref{cm}) and
(\ref{megane}) is minimized numerically with respect to the
variational parameters $\Sigma=\sigma/a^{*}_{B}$  and $\alpha=r_{\rm
\scriptscriptstyle WS}/r_{s}a^{*}_{B}$. The equilibrium values of
these parameters are shown in Figure \ref{fig1} as functions of $r_{s}$ for two
values of the magnetic field. The ratio ${\bar \Sigma}/r_{s}{\bar
\alpha}= (\sigma/r_{\rm
\scriptscriptstyle WS})_{\rm \scriptscriptstyle eq}$ should be
appreciably smaller than unity for internal self-consistency and the
extent to which this consistency criterion is satisfied is shown in
Figure \ref{fig2}. It is evident from Figures \ref{fig1} and
\ref{fig2} that for these values of
the field the harmonic approximation is applicable for the estimation
of the variational parameters whenever the
consistency criterion ${\bar \Sigma}/r_{s}{\bar
\alpha}\ll 1$ is satisfied.

\begin{figure}[h!]
\centerline{\mbox{\psfig{figure=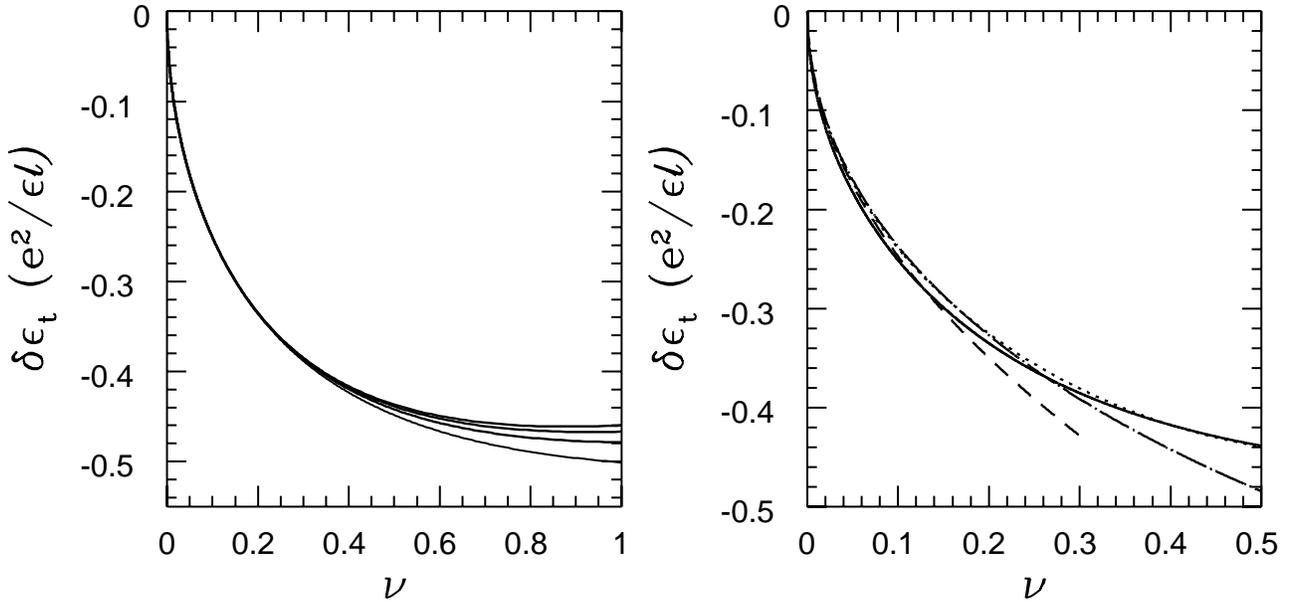, angle =0, width = 18 cm}}}
\caption{Left panel: Ground-state energy $\delta \epsilon_{t}$ of the
unpaired WC, in units of $e^{2}/\epsilon \ell$, as a function of the
filling factor $\nu$ in the range $0<\nu\leq 1$ for values of $B=5,
10, 15$ and $20$ Tesla (from bottom to top). Right panel: Ground-state
energy of the unpaired WC obtained in the present approach (full line)
compared with the CWC result of Lam and Girvin (dotted
line), with the asymptotic classical result (dashed line) and with the
energy of the Laughlin liquid (dot-dashed line, from Levesque {\it et
al.}).}
\label{fig3}
\end{figure}

Figure \ref{fig3} reports our results for the ground-state energy $\delta
\epsilon_{t}$ as a function of the filling factor in the lowest LL,
after subtraction of the kinetic energy $2 \lambda^{1/2}_B\,\, Ryd^{*}$ from
the total energy $\epsilon_{t}$ and rescaling to $e^{2}/\epsilon \ell$
energy units. The left panel in Figure \ref{fig3} shows that $\delta
\epsilon_{t}$ in these units is still weakly dependent on the field
intensity at values of $\nu$ larger than about $0.4$. The right panel
in Figure \ref{fig3} compares our results for  $\delta
\epsilon_{t}$ in the range $0<\nu \leq 0.5$ with those obtained in the
correlated Wigner crystal approach of Lam and Girvin\cite{girvin} (see
Eq. (\ref{CWC})\,). It is also seen that the classical limit of Bonsall
and Maradudin\cite{bonsa} is approximately recovered for $\nu
\rightarrow 0$. 

It is also seen from Figure \ref{fig3} (left panel) 
that the ground-state energy
of the unpaired WC as obtained in the present approach crosses the
energy of the Laughlin liquid as reported by Levesque {\it et
al.}\cite{mac} at $\nu \approx 0.25$. This value for the upper
critical filling factor of the liquid-solid transition agrees with
that reported by Vignale\cite{vignale} and, as discussed by this
author, there is strong experimental evidence supporting the fact that the crystal exists even at filling factors as large as $0.22$-$0.23$\cite{shay,rv}. However, again as discussed by Vignale, it appears that the liquid-solid transition is not a simple crossing of two phases occuring at a single value of $\nu$. Rather, it shows a complex reentrant behavior, with the liquid phase being stable at or near the Fractional Quantum Hall Effect fractions and the solid phase being stable in between.

We would also like to remark that the simple choice $\alpha=1$, which ensures that the Wigner-Seitz cell is electrically neutral, implies only small changes relative to the prescription proposed by Nagy\cite{nagy}. We in fact find that when we take $\alpha=1$ the critical filling factor at which the liquid-solid transition occurs shifts from $\nu \approx 0.25$ to $\nu \approx 0.23$.

Finally, we briefly comment on the issue of Landau Level Mixing (LLM),
as studied in detail by Zhu and Louie\cite{zhu} and by Price {\it et
al.}\cite{sarma}. LLM is important when the ratio between the magnetic
energy (of order $\hbar \omega_{0}$) and the Coulomb energy (of
order $e^{2}/(\epsilon \,r_{s}a^{*}_{B})$) becomes smaller then
unity, {\it i.e.} for $r_{s}<\lambda^{-1}_{B}$. This effect is
included in our approach through the use of a trial wave function
having a variational width\cite{sarma} and becomes
unimportant as $r_{s}$ increases and $\sigma_{\rm \scriptscriptstyle
eq}$ approaches the value $\sqrt{2}\,\ell$ for which Eq. (\ref{prima})
becomes the exact lowest-energy eigenfunction of the single-particle
Hamiltonian. This asymptotic behavior is clearly seen from Figure \ref{fig1}
(left panel).
\subsection{Analytic results in the harmonic approximation}  
Having assessed numerically the range of validity of the harmonic
approximation (HA) for the estimation of the model variational
parameters, we proceed to report a number of analytic results which
follow from it. The ground-state energy is given by
\begin{equation}\label{HA}
\epsilon^{\rm \scriptscriptstyle
HA}_{t}=\left(\Sigma^{-2}+\lambda_{B}\,\Sigma^{2}-\frac{4
\alpha}{r_{s}}+\frac{\Sigma^{2}}{\alpha\,r^{3}_{s}}+\frac{16\,\alpha^{3}}{3
\pi r_{s}}\right)\,Ryd^{*}\, .  
\end{equation}
Minimization of Eq. (\ref{HA}) with respect to $\Sigma$ and $\alpha$
yields the results shown in Figure \ref{fig1}.

Let us consider first the choice $\alpha=1$ ({\it i.e.} $r_{\rm
\scriptscriptstyle WS}=r_{s}a^{*}_{B}$), where the equilibrium value
of $\Sigma$ is
\begin{equation}\label{a1}
{\bar \Sigma}_{\alpha=1}(r_{s},
B)=\frac{r^{3/4}_{s}}{(1+\lambda_{B}\,r^{3}_{s})^{1/4}}\, .
\end{equation}
Thus, ${\bar \Sigma}_{\alpha=1}(r_{s},B)$ decreases with increasing
field, due to increased localization of the electron inside the
circular cell, and saturates to the value
$\lambda^{-1/4}_{B}$ 
corresponding to $\sigma_{\rm \scriptscriptstyle
eq}=(\hbar/m^{*}\omega_{0})^{1/2}$. The total energy per electron becomes
\begin{equation}\label{result}
\epsilon^{\rm \scriptscriptstyle HA}_{t}(r_{s},
B)=\left[2\,{\displaystyle
\frac{(1+\lambda_{B}r^{3}_{s})^{1/2}}{r^{3/2}_{s}}}
+ \left(\frac{16}{3
\pi}-4\right)\frac{1}{r_{s}}\right]\, Ryd^{*}
\end{equation}
and increases with the magnetic field, saturating to the value
\begin{equation}\label{18}
\epsilon^{\rm \scriptscriptstyle HA}_{t}(r_{s},
B\rightarrow \infty)=\left[2\,\lambda^{1/2}_{B}+ \left(\frac{16}{3
\pi}-4\right)\frac{1}{r_{s}}\right]\, Ryd^{*}\, .
\end{equation}
The first term in the brackets in Eq. (\ref{18}) is the cyclotron
zero-point energy $\hbar \omega_{0}$. On the other hand, in the limit
of vanishing field Eq. (\ref{result}) yields
\begin{eqnarray}\label{Bzer}
\epsilon^{\rm \scriptscriptstyle HA}_{t}(r_{s},B=0)
&\!=\!&\left[\left(\frac{16}{3
\pi}-4\right)\frac{1}{r_{s}}+ 
\frac{2}{r^{3/2}_{s}}\right]\, Ryd^{*}\\ \nonumber
&\!\simeq\!&
\left(-\frac{2.30}{r_{s}}+\frac{2}{r^{3/2}_{s}}\right)\, Ryd^{*} \, ,
\end{eqnarray}
which is the result obtained by Seidl {\it et al.}\cite{perdo} in
their ``PC model'' (the correlation energy is obtained from
Eq. (\ref{Bzer}) by subtracting from it the exchange energy term given by
$-8\sqrt{2}/(3\pi r_{s})$\,). The dependence of 
$\epsilon^{\rm \scriptscriptstyle HA}_{t}(r_{s},B=0)$ in
Eq. (\ref{Bzer}) on the density parameter $r_{s}$ is correct, but the
numerical coefficients are somewhat different from those which are
precisely known for the WC in zero field\cite{bonsa}. It may be remarked
that the expression for $\epsilon^{\rm \scriptscriptstyle HA}_{t}(r_{s},B=0)$
in 3D provides a lower bound to the energy of the WC\cite{Lieb}.

Minimization of the energy in Eq. (\ref{HA}) with respect to $\alpha$
is intended to approximately correct for the fact that the electrical
potential in 2D does not vanish outside the circular cell and is
expected to yield a variational lower bound for the ground-state
energy\cite{nagy}. The equilibrium value of the reduced Gaussian width
is given by an expression similar to Eq. (\ref{a1}),
\begin{equation}\label{generale}
{\bar \Sigma}_{{\bar \alpha}}(r_{s}, B)=\left(\frac{{\bar \alpha} r^{3}_{s}}
{1+{\bar \alpha} \lambda_{B}\,r^{3}_{s}}\right)^{1/4}
\end{equation}
while ${\bar \alpha}$ converges with increasing $r_{s}$ to the value
$\sqrt{\pi}/2$ (see Figure \ref{fig1}). That is, Nagy's result\cite{nagy} $r_{\rm
\scriptscriptstyle WS}=(\sqrt{\pi}/2)r_{s} a^{*}_{B}$ remains valid in
the presence of a magnetic field at low electron density. 
The expression for
the ground-state energy becomes 
\begin{equation}\label{previous}
\epsilon^{\rm \scriptscriptstyle HA}_{t}(r_{s},
B)=\left[2 \sqrt{2}\,\,{\displaystyle
\frac{(1+\lambda_{B} \sqrt{\pi} r^{3}_{s}/2)^{1/2}}{\pi^{1/4}\,r^{3/2}_{s}}}
- \frac{4 \sqrt{\pi}}{3}
\frac{1}{r_{s}}\right]\, Ryd^{*}\, .
\end{equation}
However, this simple analytic expression for the energy of the
unpaired WC in a magnetic field is not sufficiently accurate on the
energy scale needed for the comparisons made in Figure \ref{fig3}.
\section{Motion of an electron pair in a magnetic field}\label{sec3}
In this section we motivate the introduction of a paired state for the
 WC by discussing the spin-singlet ground-state and effective
 interaction potential for two electrons moving in a plane under a
 perpendicular magnetic field. As is known from the work of
 Taut\cite{taut}, this is an example of a quasi-exactly soluble
 problem with a hidden $sl_{2}$-algebraic
 structure\cite{turbiner}: an exact solution exists for special values
 of the field. Our aim will be to use the analytic form of the wave
 function determined in such a case in order to make a reasonable {\it
 Ansatz} for the variational treatment which will be developed for the
 paired phase in the next Section.

The Hamiltonian describing the two electrons is written in terms of
the relative coordinate ${\bf r}={\bf r}_{2}-{\bf r}_{1}$, of the
center-of-mass coordinate ${\bf R}=({\bf r}_{1}+{\bf r}_{2})/2$ and
of the corresponding momenta ${\bf p}$ and ${\bf P}$ as
\begin{equation}\label{hh}
{\mathcal H}=\frac{1}{4m}\left[{\bf P}+\frac{e}{c}{\bf A}_{\rm
\scriptscriptstyle CM}({\bf 
R})\right]^2 +\frac{1}{m}\left[{\bf p}+\frac{e}{c}{\bf A}_{\rm
\scriptstyle rel}({\bf 
r})\right]^2 +\frac{e^2}{r}+{\mathcal H}_{spin}
\end{equation}
where ${\bf A}_{\rm \scriptscriptstyle CM}({\bf R})={\bf B} 
\times {\bf R}$, ${\bf A}_{\rm \scriptstyle rel}({\bf r})
={\bf B}\times {\bf r}/4$ and  ${\mathcal
H}_{spin}=-(g e\hbar/m c)({\bf s}_{1}+{\bf s}_{2})$ with ${\bf
s}_{i}$ being the spins. The addition of a harmonic confining potential
merely shifts the value of the cyclotron frequency\cite{taut}.
The center-of-mass motion in the ground-state
is described by a Gaussian wave function having width given by the
magnetic length $\ell=(\hbar c/eB)^{1/2}$. The wave function
$\phi_{\rm \scriptstyle rel}({\bf r})$ for the relative motion in the
spin-singlet ground-state is even under space inversion. Following
the treatment developed by Taut\cite{taut}, an analytic expression is
obtained for the state of zero relative angular momentum ($M_{\rm
\scriptstyle rel}=0$) if the magnetic field satisfies the condition
$\gamma=1$, where $\gamma\equiv \ell_{\rm \scriptstyle rel}/a_{B}$
with $\ell_{\rm \scriptstyle rel}=\sqrt{2}\ell$ and $a_{B}=\hbar^{2}/
m e^{2}$. This state lies at relative energy $2\hbar \omega_{0}$ and
apart from a normalization factor is
\begin{equation}
\phi_{\rm \scriptstyle rel}({\bf r}) \propto \,(1+x)\,e^{-x^2/4}
\end{equation}   
with $x=r/\ell_{\rm \scriptstyle rel}$. This is, in fact, the solution
which corresponds to the lowest value of $\gamma$ and to the lowest energy.

Two main points of this two-electron problem need emphasizing. 
Firstly, the effective potential 
$V_{\rm \scriptstyle eff}(r)$ entering the Schr\" odinger 
equation for the reduced wave function 
$f_{\rm \scriptstyle eff}({\bf r})=\phi_{\rm \scriptstyle rel}({\bf r})\,\sqrt{r}$ at 
$M_{\rm \scriptstyle rel}=0$ is
\begin{equation}
V_{\rm \scriptstyle
eff}(r)=\frac{1}{4}x^{2}+\frac{\gamma}{x}-\frac{1}{4 x^{2}}
\end{equation}
and develops a minimum for $\gamma>(16/27)^{1/4}$ (see Figure
 \ref{fig4}). This fact will motivate the introduction of the paired
 phase in the next Section. Secondly, in a solid semiconducting
 medium the parameter $\gamma$ becomes
\begin{equation}
\gamma^{*}=\frac{\ell_{\rm \scriptstyle
rel}}{a^{*}_{B}}=\frac{m^{*}}{\epsilon\, m}\,\gamma\, ,
\end{equation}
$\gamma$ being the vacuum value. Thus, using 
$\ell_{\rm \scriptstyle rel}\simeq 363.5\, (B/{\rm Tesla})^{-1/2}$
\AA\,\, and the values $\epsilon\simeq 12.4$ and $m^{*}/m=0.067$ for a GaAs
heterostructure, we find that the minimum is present in $V_{\rm \scriptstyle
eff}(r)$ for all values of magnetic field {\it lower} than $B_{\rm
\scriptstyle crit}\simeq 20$ Tesla.
\section{The paired phase in a magnetic field}\label{sec4}
We return to the circular cell model to develop a variational approach to
 the total energy in the case of a localized paired phase in a
 spin-singlet configuration. Two electrons are placed inside each disk
 of radius  $r_{\rm \scriptscriptstyle WS}$, the pair wave function
 being taken in the form 
\begin{equation}\label{tr}
\psi_{\kappa, \sigma}({\bf r}_{1}, {\bf r}_{2})={\mathcal
A}_{\kappa, \sigma}\left(1+\kappa\,
\frac{|{\bf r}_{1}-{\bf r}_{2}|}{\sigma}\right)
\exp{\left(-\frac{r^2_{1}+r^2_{2}}{ 4 \sigma^{2}}\right)}\, .
\end{equation}
where $\kappa$ and $\sigma$ are variational parameters and ${\mathcal
A}_{\kappa, \sigma}$ is a normalization constant. The single-pair case
treated in Section \ref{sec3} is recovered by setting $\sigma=\ell$
and $\kappa=1/\sqrt{2}$. Notice that the relative motion of the two
electrons is described in Eq. (\ref{tr}) by a Gaussian factor times a
linear superposition of the Hermite polynomials of zeroth and first
order: thus the {\it Ansatz} (\ref{tr}) allows level mixing due to the
electron-electron interactions and is the natural extension of the
variational-width method used for the unpaired phase in Section
\ref{sec2}. More refined wave function would include higher-order
Hermite polynomials in the mixing, weighted by additional variational
parameter. 

The main properties of the wave function (\ref{tr}) are as
follows. Firstly, normalization to two electrons per cell yields
\begin{equation}
{\mathcal A}_{\kappa,\sigma}=\left[\sqrt{2}\,\pi\,
\sigma^{2}\left(1+2\,\kappa\,\sqrt{\pi}+4    
\kappa^{2}\right)^{1/2}\right]^{-1} \, .  
\end{equation}
Secondly, the most probable value of the relative distance between the
two electrons, as obtained from the appropriate maximum of the square
of the pair wave function, is
\begin{equation}\label{28}
|{\bf r}_{2}-{\bf r}_{1}|_{mp}= 
\left(\sqrt{1+16\,\kappa^2}-1\right)\,\frac{\sigma}{2\, \kappa}\, .
\end{equation}
This increases with $\kappa$ and saturates to the value
$2\sigma$. Thirdly, the single-particle density 
$\rho_{\kappa, \sigma}(r)
=\int d^{2}r'\left|
\psi_{\kappa, \sigma}({\bf r}, {\bf
r}')\right|^{2}$ can be obtained in closed form (see Appendix), with
the result
\begin{equation}\label{formulazza}
\rho_{\kappa,\sigma}(r)
=\pi \sigma^{2}\, {\mathcal A}^{2}_{\kappa,\sigma}\, 
e^{-3 r^{2}/4\sigma^{2}}\left\{
2\left(1+2\kappa^{2}+\kappa^{2}\frac{r^{2}}{\sigma^{2}}\right)       
e^{r^{2}/4\sigma^{2}}+\kappa\,\sqrt{2\pi}\,\left[
(2+\frac{r^{2}}{\sigma^{2}})\,
{\rm I}_{0}(\frac{r^{2}}{4\sigma^{2}})+\frac{r^{2}}{\sigma^{2}}\,\,
{\rm I}_{1}(\frac{r^{2}}{4\sigma^{2}})\right] \right\}\, .  
\end{equation}
This expression tends to the correct value 
$\rho_{\kappa=0,\sigma}(r)=(\pi\,\sigma^{2})^{-1}\,\,
\exp{(-r^{2}/2\sigma^{2})}$ for $\kappa \rightarrow 0$.

We proceed to evaluate the total energy per electron associated with
the wave function (\ref{tr}). We have
\begin{equation}\label{30}
\epsilon_{t}=\frac{1}{2}(\epsilon_{eb}+\epsilon_{bb}+\epsilon_{\rm
\scriptstyle pair})\, ,
\end{equation}
where for the background-background term we can use the result in
Eq. (\ref{ture}). The electron-background term is evaluated, as in the
calculation performed in Section \ref{sec2}.A, from the electrical
potential $V_{\kappa,\sigma}(r)$ created by the electron distribution 
$\rho_{\kappa,\sigma}(r)$ in Eq. (\ref{formulazza}) according to
\begin{eqnarray}\label{nuda}
\epsilon_{eb}&\!=\!&2\pi\,n_{s}\int_{0}^{r_{\rm \scriptscriptstyle WS}}\!r
\,dr \,V_{\kappa,\sigma}(r)\nonumber \\
&\!=\!&-\frac{4\,\alpha}{r_{s}}\,\int_{0}^{\infty}\!\frac{dx}{x}\,
{\widetilde \rho}_{\kappa,\sigma}(x)\,\,{\rm J}_{1}(r_{\rm
\scriptscriptstyle WS}\,x/\sigma)\,\, Ryd^{*} 
\end{eqnarray}
(see Appendix), where ${\rm J}_{n}(x)$ are Bessel functions. Here
\begin{eqnarray}\label{32}
{\widetilde \rho}_{\kappa,\sigma}(x)&\!=\!&
\frac{1}{2}\,\sum_{i=1}^{2}\langle\psi_{\kappa,
\sigma}|\,e^{-i\,{\bf x}\cdot{\bf r}_{i}/\sigma}|
\psi_{\kappa,\sigma}\rangle \nonumber \\
&\!=\!&\frac{2\,\,e^{-x^{2}/2}}
{1+2\kappa\sqrt{\pi}+4\kappa^2}
\,\,\left\{1+4\kappa^2-\kappa^2\,x^{2}
-\frac{\sqrt{\pi}}{2}\, \kappa \, e^{x^{2}/8}
\,\left[(x^{2}-4){\rm I}_{0}(x^{2}/8)-x^{2}{\rm I}_{1}
(x^{2}/8)\right]\right\}\, .
\end{eqnarray}
In the limit of strong localization ($\sigma \ll r_{\rm
\scriptscriptstyle WS}$), we obtain the harmonic-approximation result
\begin{equation}\label{pippo1}
\epsilon^{\rm \scriptscriptstyle HA}_{eb}=
\left[-\frac{8\,\alpha}{r_{s}}
+\frac{2\,\Sigma^{2}}{\alpha \,r^3_{s}}\,
\frac{2+5\kappa\sqrt{\pi}+12\kappa^2}{1+2\kappa\sqrt{\pi}+4\kappa^2} 
\right]\,\, Ryd^{*}\, . 
\end{equation} 
The same limiting result can be obtained from the electrical
potential $V_{\mbox{\tiny in}}(r)$ created by the background disk,
which is still given by Eq. (\ref{vale}).

Finally the contribution $\epsilon_{\rm
\scriptstyle pair}$  in Eq. (\ref{30}) is given by 
\begin{eqnarray}
\epsilon_{\rm \scriptstyle pair}&\!=\!&\frac{1}{2}\,\langle
\psi_{\kappa, \sigma}|{\mathcal
H}|\psi_{\kappa, \sigma}\rangle\nonumber \\
&\!=\!& \frac{1}{2}\,{\mathcal
A}^{2}_{\kappa, \sigma}\Big(\langle\phi_{\rm \scriptstyle rel}|\phi_{\rm
\scriptstyle rel}\rangle\langle\phi_{\rm \scriptscriptstyle
CM}|{\mathcal H}_{\rm \scriptscriptstyle CM}|\phi_{\rm \scriptscriptstyle
CM}\rangle + \langle\phi_{\rm \scriptscriptstyle CM}|\phi_{\rm
\scriptscriptstyle CM}\rangle\langle\phi_{\rm \scriptstyle
rel}|{\mathcal H}_{\rm \scriptstyle rel}|\phi_{\rm \scriptstyle
rel}\rangle\Big)\, ,
\end{eqnarray}
the center-of-mass and relative motion states and Hamiltonians being
immediately obtained from Eqs. (\ref{hh}) and (\ref{tr}). The result
is
\begin{equation}\label{35}
\epsilon_{\rm \scriptstyle
pair}=\left(\frac{1}{2\Sigma^{2}}\,\,\frac{2+3\,\kappa\,\sqrt{\pi}
+8\kappa^{2}}{1+2\kappa\,\sqrt{\pi}+4\kappa^{2}}
+2\lambda_{B}\Sigma^{2}\,\,\frac{2+5\kappa\, 
\sqrt{\pi}+12\kappa^{2}}{1+2\kappa\,\sqrt{\pi}+4\kappa^{2}}
+\frac{\sqrt{\pi}}{\Sigma}\frac{1+4\kappa/\sqrt{\pi}+2\kappa^2}
{1+2\sqrt{\pi}\kappa+4\kappa^{2}} \right)\,\,Ryd^{*}\, . 
\end{equation}
The last term in this equation arises from the electron-electron
interaction. 
\section{Total energy of the paired phase}\label{sec5}
\begin{figure}[h!]
\centerline{\mbox{\psfig{figure=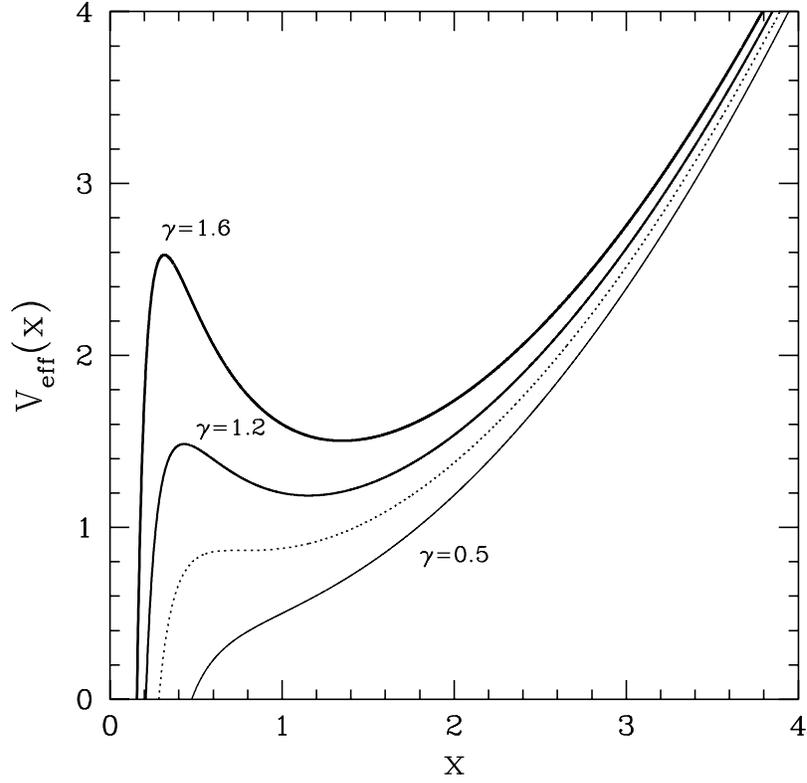, angle =0, width = 12 cm}}}
\caption{Effective potential $V_{\rm \scriptstyle
eff}(r)$ for the relative motion at zero relative angular
momentum. The dotted line corresponds to the critical value
$\gamma_{\rm \scriptstyle crit}=(16/27)^{1/4}$, above which a minimum is
present. The dotted line has an inflection point at $x_{\rm
\scriptstyle crit}=2/(3\gamma_{\rm \scriptstyle crit})$.} 
\label{fig4}
\end{figure}

\begin{figure}[h!]
\centerline{\mbox{\psfig{figure=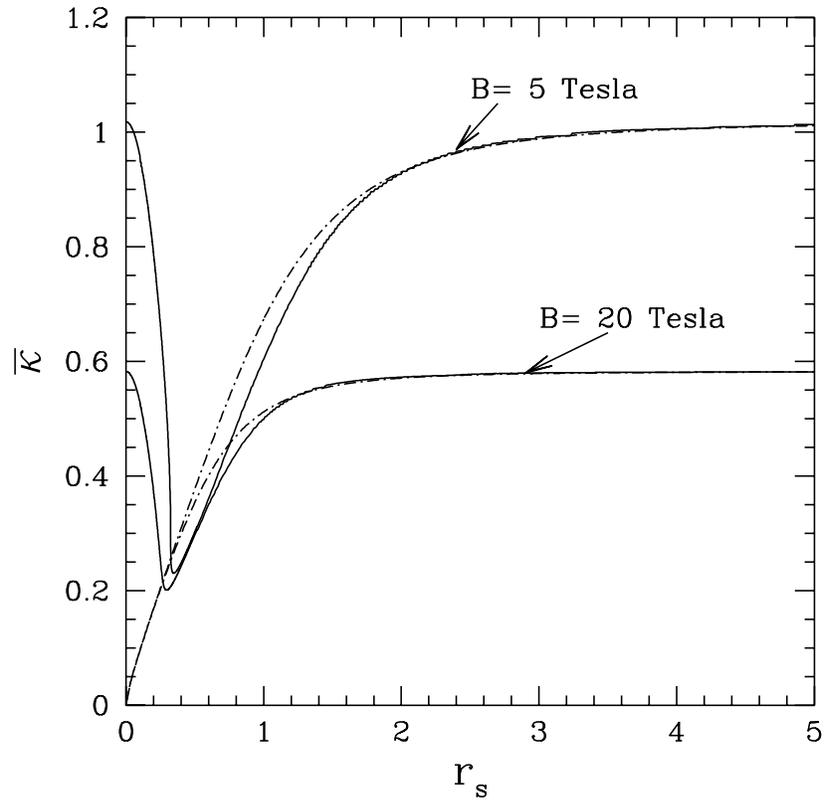, angle =0, width = 12 cm}}}
\caption{The parameter ${\bar \kappa}$ as a function of $r_{s}$ 
for two values of the field $B$ in the paired phase (full lines). The
dash-dotted lines show the results of the harmonic approximation.}
\label{fig5}
\end{figure}

The total variational energy per electron in the paired circular-cell
approximation is obtained from Eqs. (\ref{30}), (\ref{nuda}), (\ref{35}) and
(\ref{ture}). It depends on the three variational parameters
$\Sigma=\sigma/a^{*}_{B}$, $\alpha=r_{\rm \scriptscriptstyle
WS}/(r_{s}a^{*}_{B})$ and $\kappa$. The equilibrium values of $\Sigma$
and $\alpha$ as functions of $r_{s}$ show the same trends as displayed
in Figure \ref{fig1} for the unpaired phase. 
However, the asymptotic value of ${\bar \Sigma}$ for an electron pair 
is reduced by a
factor of $\sqrt{2}$ 
and that of ${\bar \alpha}$ is ${\bar \alpha}=\sqrt{\pi/2}$, 
increased by a factor $\sqrt{2}$ over the unpaired state as one
expects from the double occupancy of the cell. Internal consistency of
the theory is ensured by the ratio $(\sigma/r_{\rm \scriptscriptstyle
WS})_{\rm \scriptscriptstyle eq}$ being smaller than about $0.5$, both
in the full calculation and in the harmonic approximation, for values
of $r_{s}>1$ and of the field $B>5$ Tesla. The equilibrium value of
$\kappa$ as a function of $r_{s}$ is displayed in Figure \ref{fig4}
for two values of the field. The two electrons in each cell are pushed
closer together as the magnetic field increases, leading from
Eq. (\ref{28}) to a decrease of ${\bar \kappa}$ with increasing field as
is shown in Figure \ref{fig5}. 

\begin{figure}[h!]
\centerline{\mbox{\psfig{figure=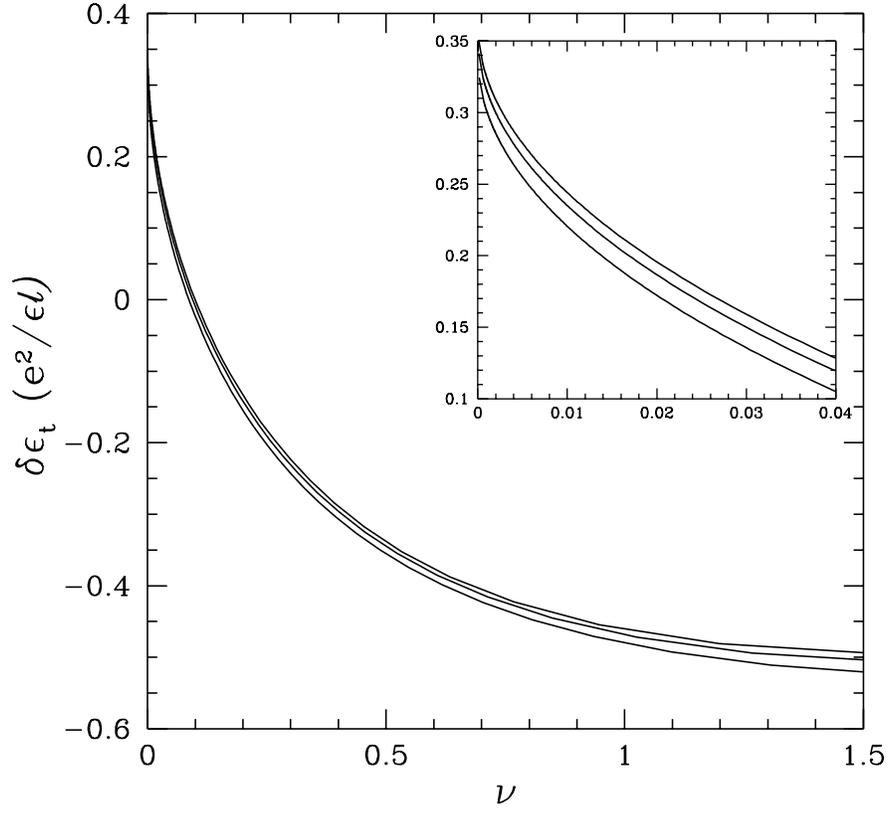, angle =0, width = 12 cm}}}
\caption{The energy $\delta \epsilon_{t}$ of the paired phase as a
function of the filling factor $\nu$ for various values of the
magnetic field (increasing from $10$ to $20$ Tesla from bottom to
top). The inset shows an 
enlarged view of the region $\nu \rightarrow 0$.}
\label{fig6}
\end{figure}

Figure \ref{fig6} reports the energy $\delta \epsilon_{t}$ of the
paired phase (that is, after subtraction of the kinetic energy in the
lowest LL and reduction to $e^{2}/(\epsilon\,\ell)$ energy units) as a
function of the filling factor extending up to $\nu=1.5$. The residual
dependence of  $\delta \epsilon_{t}$ on the magnetic field extends
into the high-$r_{s}$ (low-$\nu$) regime, as illustrated in the inset
in Figure \ref{fig5} on a magnified scale, and a finite value is
attained by  $\delta \epsilon_{t}$ as $\nu \rightarrow 0$. These
features are a consequence of the Coulomb interaction between the two
electrons inside each cell. The result in Figure \ref{fig6} have been
fitted to the functional form  $\delta \epsilon_{t}=f(\nu)\,\,
e^{2}/(\epsilon\,\ell)$, where
\begin{equation}\label{interpo}
f(\nu)=\frac{a\,\nu^{1/2}+b\,\nu^{3/2}+c\,\nu^{5/2}+d}{e\,\nu^{1/2}+f}\, .
\end{equation}
The values of the coefficients in this fitting formula are reported in
Table \ref{table1} for various values of the magnetic field.

\begin{figure}[h!]
\centerline{\mbox{\psfig{figure=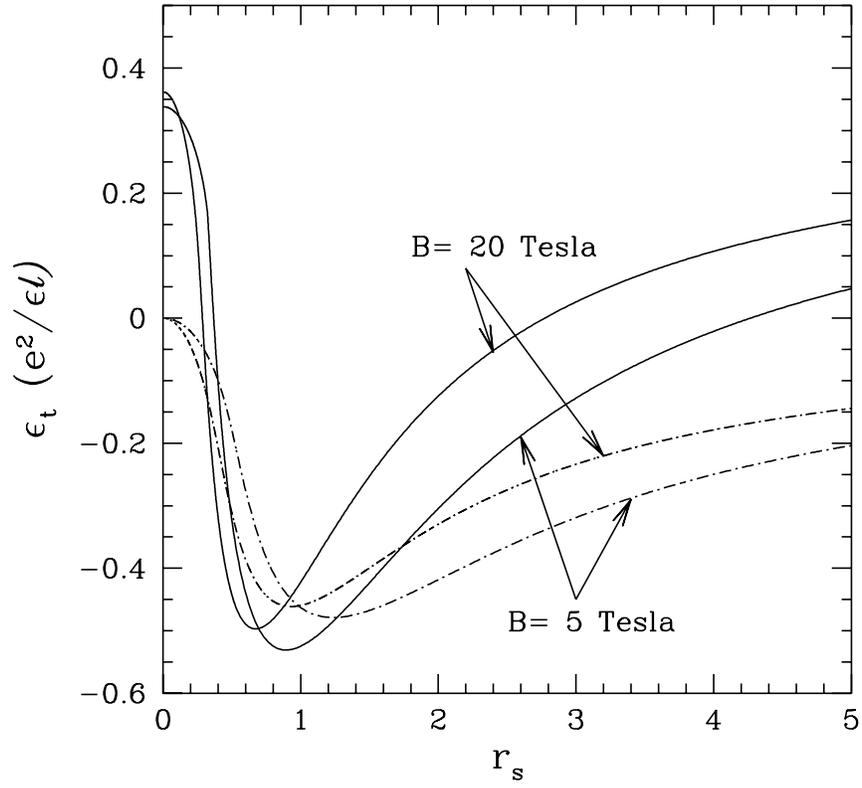, angle =0, width = 12 cm}}}
\caption{Comparison of the total energies of the paired state (full
lines) and of the unpaired state (dash-dotted lines) for two values of
the magnetic field.}
\label{fig7}
\end{figure}

Figure \ref{fig7} compares the energies per electron of the unpaired
and paired phases as functions of $r_{s}$. The two curves cross at two
values of $r_{s}$ for each value of the field, but the crossing at
lower $r_{s}$ is to be discarded as it corresponds to situations where
charge leakage out of the circular cell is unacceptably high
($(\sigma/r_{\rm \scriptscriptstyle WS})_{\rm \scriptscriptstyle
eq}\approx 1$\,). The conclusion from the physically significant
crossing at larger $r_{s}$, therefore, is that at each value of the
magnetic field in the range from $5$ to $20$ Tesla the paired phase
becomes stable relative to the unpaired one as $r_{s}$ decreases.

From the location of the physically acceptable crossing of the
energies of the two phases in Figure \ref{fig7} we obtain their phase
boundary in the $(r_{s}, \lambda_{B})$ plane, which is reported in
Figure \ref{fig8}. This shows a region of thermodynamic stability for
the paired phase (at least relative to the unpaired one) in the left
portion of the plane, before charge leakage outside the cell boundary
is expected to occur in an important way so that the model becomes
unreliable (to the left of the dash-dotted line in the Figure). Within 
numerical accuracy we find that the boundary between the two phases 
is actually set by the $\nu=1$ line in
the plane\cite{25}. Therefore, the present simple model predicts that
electron clusterization may occur only in Landau levels above the
lowest one. Of course, the unpaired state in the $\nu <1$ region (on
the right of the $\nu=1$ line) could become stable against the
Laughlin liquid only for $\nu<0.25$, as already discussed in
connection with Figure \ref{fig3}.

We should emphasize that the ground-state energy difference that we
estimate between the paired and the unpaired state is quite
appreciable as long as the magnetic field intensity lies in the range
that we have illustrated in our calculations. The difference decreases
with the field intensity, so that more refined calculations would be
needed at low fields.
\section{Summary and Conclusions}
We have presented a variational study which approximately accounts for
the ground-state energetics of a 2D many-electron system with Coulomb
interactions at low density and in the presence of a perpendicular
magnetic field. The method consists of an improved 2D version of the
well-known spherical cell approximation and is applied to two distinct
situations, a single-electron state and a spin-singlet paired
state. For the former it gives new analytical and numerical results
which are in close agreement with the state-of-the-art calculations on
the 2D Wigner crystal.

The evaluation of the spin-paired state has been motivated by previous
studies and suggests that this state may be stable in a range of
system parameters corresponding to Landau levels above the lowest
one. It remains to be seen whether such a paired state would be
confirmed in more sophisticated treatments allowing in particular for
inter-cell corrections. It may be also emphasized at this point that we
have not examined the stability of the spin-singlet paired state against
the emergence of spin-polarized states. Evidence from measurements of
Knight shift of the $^{71}{\rm Ga}$ nuclei in n-doped GaAs\cite{Barrett}
indicates that this quantity, which is proportional to the spin polarization, 
drops precipitously on either side of $\nu=1$, which is evidence that the charged 
excitations of the $\nu=1$ ground state are finite-size Skyrmions.
\acknowledgements
This work was partially supported by MURST through the PRIN Initiative.
One of us (M.P.) wishes to thank the Physics Department of the
University of Cyprus for their hospitality during part of this work. 
\section*{Appendix. Details of Analytic calculations}
We report in this Appendix some details on the derivation of some
analytic results given in the main text.

\subsection{Equations (\ref{cmm}) and (\ref{cm})}
The single-electron potential $V_{\rm \scriptstyle SP}(r)$ is 
\begin{equation}
V_{\rm \scriptstyle
SP}(r)=-(2\,a^{*}_{B}/\sigma)\,\int_{0}^{+\infty}\! dx\,
{\widetilde \rho}_{\rm \scriptstyle SP}(x)\,{\rm J}_{0}(x\,r/\sigma)
\end{equation}
(in $Ryd^{*}$), where
\begin{equation}
{\widetilde \rho}_{\rm \scriptstyle SP}(x)=\int 
\!d^{2}r\,\phi^{2}_{\rm \scriptstyle SP}(r)\,{\rm
J}_{0}(x\,r/\sigma)=\exp{(-x^{2}/4)}\, .
\end{equation}
Equation (\ref{cmm}) in the main text follows by using the result
\begin{equation}
\int_{0}^{+\infty}\!\exp{(-x^{2}/4)}\,{\rm
J}_{0}(a\,x)=\sqrt{\pi}\,\exp{(-a^{2}/2)}\,{\rm I}_{0}(a^{2}/2)\, .
\end{equation}
Finally, the integral in Eq. (\ref{complete}) is carried out with the
help of the relation
\begin{equation}
\int_{0}^{b}\!x\,\exp{(-x^{2}/2)}\,{\rm
I}_{0}(x^{2}/2)\,dx=\frac{b^2}{2}\,e^{-b^2/2}\,\left[{\rm
I}_{0}(b^2/2)+{\rm I}_{1}(b^2/2)\right]\, . 
\end{equation}
leading to Eq. (\ref{cm}) in the main text.
\subsection{Equation (\ref{formulazza})}
The single-particle density $\rho_{\kappa, \sigma}(r)$ is calculated
from the expression
\begin{equation}
\rho_{\kappa, \sigma}(r)
=2\sigma^{2}\,{\mathcal A}^{2}_{\kappa,\sigma}\,  
\exp{(-r^{2}/\sigma^{2})}\,\int_{0}^{2 \pi}\!d\theta 
\int_{0}^{+\infty}\!x\,dx\,(1+\kappa
x)^{2}\,\exp{(-x\,r\cos{\theta}/\sigma-x^{2}/2)}\, .
\end{equation}
Eq. (\ref{formulazza}) is obtained by using the results
\begin{equation}
\int_{0}^{2\pi}\exp{(-xy\cos{\theta})}\,d\theta=
2 \pi\, {\rm I}_{0}(xy)
\end{equation}
and
\begin{eqnarray}
&&\int_{0}^{+\infty}x\,(1+\kappa
x)^{2}\,\exp{(-x^{2}/2)}\,{\rm I}_{0}(xy)
\,dx =\nonumber \\
&\!=\!&\,\exp{(y^{2}/4)}
\left\{\left(1+2\kappa^{2}+\kappa^{2}y^{2}\right)  
\exp{(y^{2}/4)}+\kappa\sqrt{\pi/2}\left[ (2+y^{2})\,\sqrt{\pi/2}\,
{\rm I}_{0}(y^{2}/4)+y^{2}\,
{\rm I}_{1}(y^{2}/4)\right]\right\}\, .
\end{eqnarray}
\subsection{Equation (\ref{nuda})}
The electrical potential $V_{\rm \scriptstyle \kappa, \sigma}(r)$
created by the electron distribution $\rho_{\kappa,\sigma}(r)$ in
Eq. (\ref{formulazza}) is written as
\begin{equation}
V_{\kappa,\sigma}(r)=-(2a^{*}_{B}/\sigma)\,\int_{0}^{+\infty}\! dx\,
\,{\widetilde \rho}_{\kappa,\sigma}(x){\rm J}_{0}(x\,r/\sigma)
\end{equation}
(in $Ryd^{*}$), where in essence ${\widetilde
\rho}_{\kappa,\sigma}(x)$ is the Fourier transform of the electron
density distribution (see Eq. (\ref{32})\,). After inserting this
formula into the first line of Eq. (\ref{nuda}) and interchanging the
order of the two integrations, the integration over $r$ can be carried
out with the help of the result
\begin{equation}
\int_{0}^{b}{\rm J}_{0}(ax)\,x\, dx=\frac{b}{a}\,\,{\rm J}_{1}(a\,b)\, .
\end{equation}
This yields the second line of Eq. (\ref{nuda}) in the main text.

\newpage
\begin{table}
\begin{center}
\caption{Coefficients of the interpolation formula in Eq. (\ref{interpo}).}
\label{table1}
\begin{tabular}{|c|c|c|c|c|c|c|}
B (Tesla) & {\emph a} & {\emph b}& {\emph c} & {\emph d} & {\emph e} 
&  {\emph f} \\ \hline
 5 &  -0.6621  &  0.2769   &  -0.0647  &
 0.1751 &  -0.0072  &  0.5578 \\ 
 10 &  -0.6435  &  0.2715  &  -0.0588 &  0.1828 &
 -0.0231  &  0.5406 \\ 
 15 &  -0.6325  &  0.2655  &  -0.0548 &  0.1869 &
 -0.0288  &  0.5302 \\ 
 20 &  -0.6255  &  0.2612  &  -0.0522  &  0.1894 &  -0.0315  &  0.5237
\end{tabular}
\end{center}
\end{table}
\begin{figure}[h!]
\centerline{\mbox{\psfig{figure=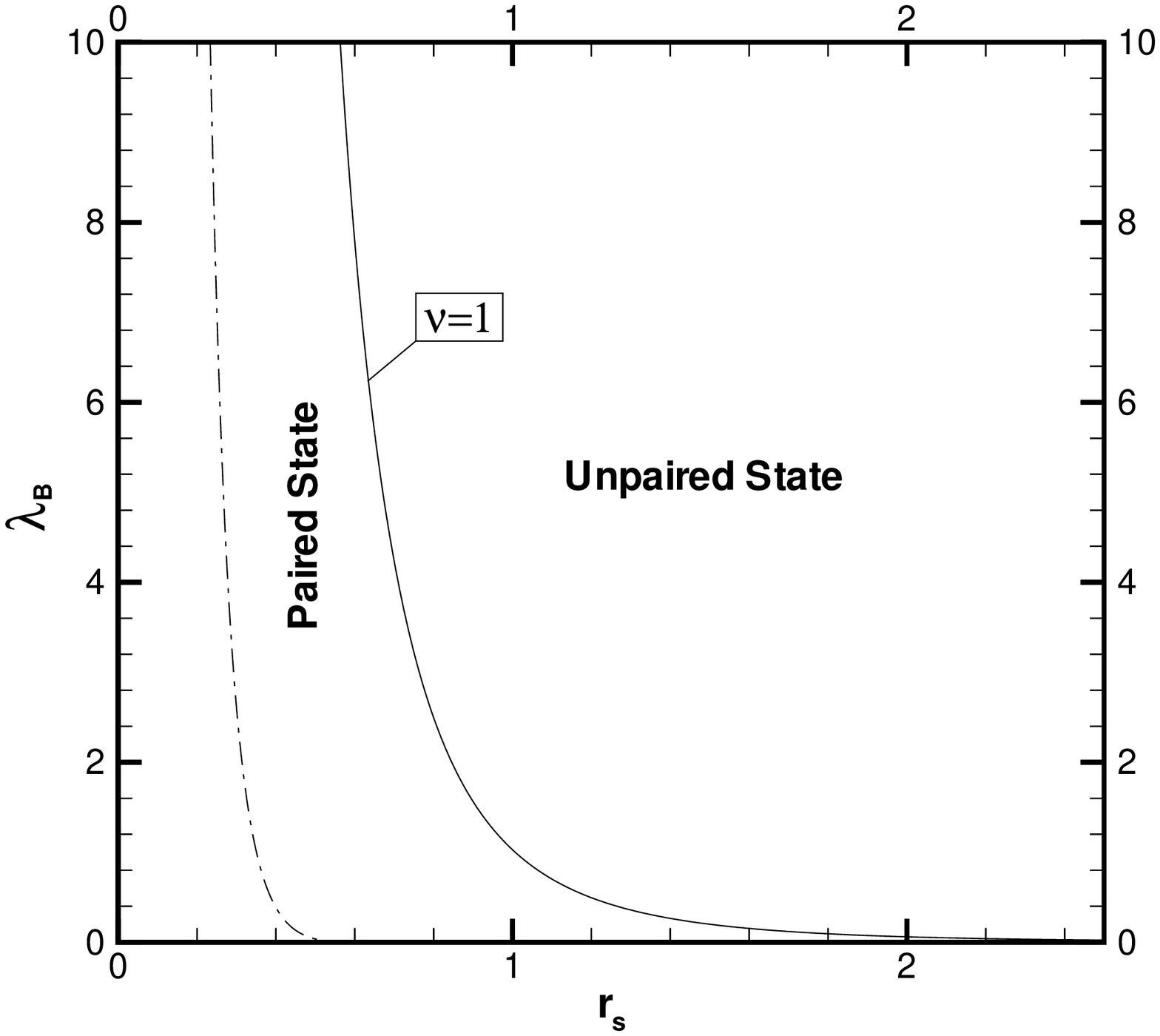, angle =0, width = 12 cm}}}
\caption{Boundary between the singlet-spin paired state and the
unpaired state in the $(r_{s}, \lambda_{B})$ plane. The boundary is
set by the $\nu=1$ line. The dash-dotted line shows where the
consistency ratio ${\bar \Sigma}/r_{s}{\bar \alpha}$ for the paired
state becomes unity.}
\label{fig8}
\end{figure}

\begin{references}
\bibitem{prange}
See {\it e.g.} {\it The Quantum Hall Effect},
eds. R. E. Prange and S. M. Girvin  (Springer, New York 1987); 
{\it Perspectives in Quantum Hall Effects},
eds. S. Das Sarma and  A. Pinczuk (Wiley, New York 1997).
\bibitem{shay}
M. Shayegan, in {\it Perspectives in Quantum Hall Effects},
eds. S. Das Sarma and  A. Pinczuk (Wiley, New York 1997), p. 343.
\bibitem{jim}
M. P. Lilly, K. B. Cooper, J. P. Eisenstein, L. N. Pfeiffer, and
K. W. West, Phys. Rev. Lett. {\bf 82}, 394 (1999); R. R. Du,
D. C. Tsui, H. L. Stormer, L. N. Pfeiffer, K. W. Baldwin, and
K. W. West, Solid State Commun. {\bf 109}, 389 (1999); K. B. Cooper,
M. P. Lilly, J. P. Eisenstein, L. N. Pfeiffer, and
K. W. West, Phys. Rev. B {\bf 60}, R11285 (1999).
\bibitem{KFS}
A. A. Koulakov, M. M. Fogler, and B. I. Shklovskii,
Phys. Rev. Lett. {\bf 76}, 499 (1996); M. M. Fogler, A. A. Koulakov
and B. I. Shklovskii, Phys. Rev. B {\bf 54}, 1853 (1996); R. Moessner
and J. T. Chalker, Phys. Rev. B {\bf 54}, 5006 (1996); M. M. Fogler
and A. A. Koulakov, Phys. Rev. B {\bf 55}, 9326 (1997).
\bibitem{haldane}
F. D. M. Haldane, E. H. Rezayi, and Kun Yang, 
Phys. Rev. Lett. {\bf 85}, 5396 (2000).
\bibitem{shibata}
N. Shibata and D. Yoshioka, Phys. Rev. Lett. {\bf 86}, 5755 (2001).
\bibitem{girless}
S. M. Girvin, {\it The Quantum Hall Effect: Novel Excitations and
Broken Symmetries}, Lectures at Ecole d'Et\'e Les Houches (1998). 
\bibitem{taut}
M. Taut, Phys. Rev. A {\bf 48}, 3561 (1993); 
M. Taut, J. Phys. A {\bf 27}, 1045 and 4723 (1994); 
M. Taut, J. Phys.: Condens. Matter {\bf 12}, 3689 (2000).
\bibitem{russian} G. V. Shuster and A. I. Kozinskaya, Fiz. Tverd. Tela
{\bf 13}, 1240 (1971)  [Sov. Phys. Solid State {\bf 13}, 1038 (1971)].
\bibitem{ma} K. Moulopoulos and N. W. Ashcroft, Phys. Rev. B {\bf 48}, 11646
(1993).  
\bibitem{experimental} 
G. M. Summers, R. J. Warburton, J. G. Michels,
R. J. Nicholas, J. J. Harris, and C. T. Foxon,  Phys. Rev. Lett. {\bf  70},
 2150 (1993); 
R. B. Dunford, E. E. Mitchell, R. G. Clark, V. A. Stadnik, F. F. Fang,
R. Newbury, R. H. McKenzie, R. P. Starrett, P. J. Wang, and
B. S. Meyerson, J. Phys.: Condens. Matt.  {\bf 9}, 1565 (1997). 
\bibitem{mou}
See, {\it e.g.} Appendix A in Ref. 10 for 
the 3D version of the 
cell approximation in zero magnetic field. A spherical cell is
there exploited
in both a point-charge and a distributed-charge approach, to deal with
the WC and with an orientationally symmetric paired crystalline
phase.
\bibitem{nagy}
I. Nagy, Phys. Rev. B {\bf 60}, 4404 (1999).
\bibitem{loz}
Y. E. Lozovik and V. I. Yudson, JEPT Lett. {\bf 22}, 11 (1975).
\bibitem{bonsa}
L. Bonsall and A. A. Maradudin, Phys. Rev. B {\bf 15}, 1959 (1977); 
E. Cockayne and V. Elser, Phys. Rev. B {\bf 43}, 623 (1991).
\bibitem{yoshi}
D. Yoshioka and H. Fukuyama, J. Phys. Soc. Jpn. {\bf 47}, 394 (1979); 
D. Yoshioka and P. A. Lee, Phys. Rev. B {\bf 27}, 4986 (1983); 
K. Maki and X. Zotos, Phys. Rev. B {\bf 28}, 4349 (1983).
\bibitem{mac}
D. Levesque, J. J. Weis, and A. H. MacDonald, Phys. Rev. B {\bf 30}, 
1056 (1984). 
\bibitem{girvin}
P. K. Lam and S. M. Girvin, Phys. Rev. B {\bf 30}, 473 (1984).
\bibitem{vignale}
G. Vignale, Phys. Rev. B {\bf 47}, 10105 (1993).
\bibitem{abram}
{\it Handbook of Mathematical Functions},
eds. M. Abramowitz and  I. A. Stegun (Dover, New York 1972).
\bibitem{rv}
D. C. Tsui, H. L. Stormer and A. C. Gossard, \prl {\bf 48}, 1559 (1982);
E. Y. Andrei, G. Deville, D. C. Glattli, F. I. B. Williams, E. Paris and N. Etienne, \prl {\bf 60}, 2765 (1988); R. L. Willett, H. L. Stormer, D. C. Tsui, L. N. Pfeiffer, K. W. West and K. W. Baldwin, \prb {\bf 38}, 7881 (1988); H. W. Jiang, R. L. Willett, H. L. Stormer, D. C. Tsui, L. N. Pfeiffer and K. W. West, \prl {\bf 65}, 633 (1990); H. W. Jiang, H. L. Stormer, D. C. Tsui, L. N. Pfeiffer and K. W. West, \prb {\bf 44}, 8107 (1991); V. J. Goldman, M. Santos, M. Shayegan and J. E. Cunninghan, \prl {\bf 65}, 2189 (1990); D. C. Glattli, G. Deville, V. Duburcq, F. I. B. Williams, E. Paris, N. Etienne and E. Y. Andrei, Surf. Sci. {\bf 229}, 344 (1990); H. Buhmann, W. Joss, K. von Klitzing, I. V. Kukushkin, A. S. Plaut, G. Martinez, K. Ploog and V. Timofeev, \prl {\bf 66}, 926 (1991); F. I. B. Williams, P. A. Wright, R. G. Clark, E. Y. Andrei, G. Deville, D. C. Glattli, O. Probst, B. Etienne, C. Dorin, C. T. Foxon and J. J. Harris, \prl {\bf 66}, 3285 (1991); R. L. Willett, M. A. Paalanen, R. R. Ruel, K. W. West, L. N. Pfeiffer and D. J. Bishop, \prl {\bf 65}, 112 (1990).
\bibitem{zhu}
X. Zhu and S. G. Louie, Phys. Rev. Lett. {\bf 70}, 335 (1993).
\bibitem{sarma}
R. Price, X. Zhu, S. Das Sarma, and P. M. Platzman, Phys. Rev. B {\bf 51},
2017 (1995).
\bibitem{perdo}
M. Seidl, J. P. Perdew, and M. Levy, Phys. Rev. A {\bf 59}, 51 (1999); 
M. Seidl and J. P. Perdew, Phys. Rev. B {\bf 50}, 5744 (1994).
\bibitem{Lieb} 
E. H. Lieb and H. Narnhofer, J. Stat. Phys. {\bf 12}, 291 (1975).
\bibitem{turbiner} A. Turbiner, Phys. Rev. A {\bf 50}, 5335 (1994).
\bibitem{25} 
The lines of constant $\nu$ in the $(r_{s}, \lambda_{B})$ plane are
described by the equation $\lambda_{B}=1/(\nu^{2}r^{4}_{s})$, so that
the line $\nu=1$ in Figure \ref{fig8} separate the lowest LL (on its
right) from the higher ones (on its left).
\bibitem{last}
S. L. Sondhi, A. Karlhede, S. A. Kivelson and E. H. Rezayi, \prb {\bf 47}, 16419 (1993).
\bibitem{Barrett}
S. E. Barrett, G. Dabbagh, L. N. Pfeiffer, K. W. West, and R. Tycko, Phys. Rev. Lett.
{\bf 74}, 5112 (1995).
\end{references}
\end{document}